\documentclass[conference]{IEEEtran}
\usepackage{graphicx}
\usepackage{amsmath}
\usepackage{array}
\usepackage{enumitem}
\usepackage{algorithm}
\usepackage{algpseudocode}
\usepackage{amsmath}
\usepackage{color,soul}

\usepackage{amssymb}    
\usepackage{pifont}      
\usepackage{array}       
\usepackage{booktabs}    
\usepackage{caption}     

\newcommand{\xmark}{\ding{55}}

\begin{document}

\title{Machine Learning Driven Smishing Detection Framework for Mobile Security}

\author{
\IEEEauthorblockN{
Diksha Goel$^{1,2^*}$\thanks{Corresponding author}, 
Hussain Ahmad$^{3}$, 
Ankit Kumar Jain$^{2}$, 
Nikhil Kumar Goel$^{4}$
}
\IEEEauthorblockA{$^{1}$CSIRO's Data61, Australia}
\IEEEauthorblockA{$^{2}$National Institute of Technology, Kurukshetra, India}
\IEEEauthorblockA{$^{3}$University of Adelaide, Australia}
\IEEEauthorblockA{$^{4}${PGIMS, Haryana}, India}
Email: diksha.goel@data61.csiro.au; hussain.ahmad@adelaide.edu.au; ankitjain@nitkkr.ac.in; nikhilgoel.kkr@gmail.com
}

\maketitle

\begin{abstract}
The increasing reliance on smartphones for communication, financial transactions, and personal data management has made them prime targets for cyberattacks, particularly smishing, a sophisticated variant of phishing conducted via SMS. Despite the growing threat, traditional detection methods often struggle with the informal and evolving nature of SMS language, which includes abbreviations, slang, and short forms. This paper presents an enhanced content-based smishing detection framework that leverages advanced text normalization techniques to improve detection accuracy. By converting non-standard text into its standardized form, the proposed model enhances the efficacy of machine learning classifiers, particularly the Naïve Bayesian classifier, in distinguishing smishing messages from legitimate ones. Our experimental results, validated on a publicly available dataset, demonstrate a detection accuracy of 96.2\%, with a low False Positive Rate of 3.87\% and False Negative Rate of 2.85\%. This approach significantly outperforms existing methodologies, providing a robust solution to the increasingly sophisticated threat of smishing in the mobile environment.

\end{abstract}


\section{Introduction}

The rapid advancement of technology has dramatically transformed our operational frameworks, mainly through the proliferation of smart devices such as smartphones and tablets \cite{Mishra2019, ahmad2024smart}. Among these, smartphones have become an essential part of our daily lives, offering unparalleled convenience and functionality \cite{seo2024, akande2023}. With over {7.211 billion users worldwide and 23 billion texts} messages sent daily, smartphones are now indispensable tools for communication, business, and personal tasks. This widespread adoption, however, has come with its drawbacks. The increasing reliance on smartphones has made them prime targets for cybercriminals, leading to a significant rise in cybercrime incidents \cite{verkijika2018understanding, jain2020predicting}\footnotetext{* Corresponding Author}.

As the integration of smartphones into our lives deepens, so does the sophistication of cyber threats targeting these devices \cite{mishra2020smishing}. Cyber threats targeting smartphones have evolved in both sophistication and frequency. These threats include viruses, malware, Denial of Service (DoS) attacks \cite{somani2016ddos, ahmad2023}, and social engineering tactics like phishing attacks \cite{mouton2016social}. Among these, phishing has emerged as one of the most pervasive and damaging forms of cyberattack \cite{Mishra2019}, where attackers impersonate legitimate entities to deceive users into divulging sensitive information such as credit card numbers, bank account details, and personal data \cite{seo2024}. These attacks often result in substantial financial losses for both individuals and organizations \cite{goel2017mobile}.

Building upon the persistent nature of phishing, this threat has evolved significantly, particularly impacting mobile devices \cite{kohilan2023}. Since the term was first coined in 1996, phishing has continuously adapted to exploit new technological vulnerabilities \cite{jain2017phishing}. In recent years, phishing attacks have increasingly targeted mobile devices, leveraging the ubiquitous presence of smartphones \cite{mishra2020smishing, ahmad2024pro}. Attackers now commonly employ Trojans, viruses, and ransomware in conjunction with phishing techniques to compromise smartphone security \cite{brewer2016ransomware, jayalath2024}.

The immediacy and personal nature of mobile communication make smartphones particularly susceptible to phishing attacks \cite{akande2023}. This vulnerability has led to the emergence of \textbf{\textit{Smishing}} attack, a portmanteau of smartphones and phishing, which explicitly targets users through SMS (Short Message Service) text messages \cite{jain12024s, chopra2024chatnvd}. Smishing attacks leverage text messages to trick users into clicking malicious links or providing confidential information, exploiting the trust users place in familiar communication channels \cite{joo2017s, goel2023enhancing}. Smishing highlights the unique vulnerabilities inherent in SMS communication \cite{kohilan2023, goel2018smishing}. 

Research shows that SMS messages have an exceptionally high open rate, averaging 98\%, almost five times greater than email. Additionally, 90\% of people open a text within three minutes of receiving it, highlighting SMS's immediacy and effectiveness \cite{goel2018overview}. Preferences also lean toward SMS, with 85\% of smartphone users favouring it over emails or calls \cite{sender2024sms}. The sheer scale of smartphone usage, growing from 1.57 billion users in 2014 to an estimated 2.87 billion by 2020 \cite{statista2020smartphone}, provides a vast target base for attackers. Despite the high engagement rates, awareness of smishing remains low, with a 2017 survey revealing that 45\% of users had received smishing messages, yet only 16\% could correctly identify the threat \cite{wombat2018state}. By combining the reach and personal touch of SMS with deceptive phishing tactics, smishing has become a particularly effective and dangerous method for cybercriminals to compromise smartphone security \cite{awumee2023, abdulsatar2024}.

The real-world implications of smishing further underscore the critical need for effective countermeasures. The impact of smishing is not just theoretical; real-world cases demonstrate the severe consequences of these attacks. For instance, in 2016, a client of UK bank Santander lost £22,700 in a smishing scam \cite{infosec2018smishing}. The frequency of these attacks, coupled with the lack of user awareness and the high success rate of smishing, underscores the urgent need for more effective detection and prevention mechanisms \cite{mishra2020smishing, awumee2023}.

Despite the apparent necessity for robust smishing detection mechanisms, existing solutions need to be more robust in adequately addressing the nuanced nature of smishing threats. Traditional smishing detection methods predominantly rely on signature-based and rule-based approaches, which need to be revised in their ability to adapt to the evolving tactics employed by cybercriminals. These conventional solutions typically focus on identifying known malicious patterns or predefined keywords within SMS content. However, this strategy could be more effective against sophisticated smishing campaigns that employ dynamic and evasive techniques to circumvent detection. One of the primary limitations of current solutions is their inadequate handling of the informal language, slang, and abbreviations commonly used in SMS communication. Unlike formal emails or web content, SMS messages often contain creative misspellings, intentional grammatical errors, and localized jargon that obscure malicious intent. For example, a smishing attempt might use abbreviations like "ur" instead of "your" or incorporate slang terms such as "lol" to make the message appear more legitimate and relatable. These linguistic nuances pose a significant challenge for traditional spam filters, which typically need to be designed to interpret or normalize such variations effectively.

Furthermore, the contextual and semantic variability inherent in SMS language exacerbates the difficulty of accurately identifying smishing attempts. Machine learning models employed by existing solutions often need help to discern the intent behind messages that mix legitimate content with deceptive elements. Another critical issue is the rapid evolution of smishing tactics, where attackers continuously refine their methods to exploit new vulnerabilities and bypass existing defences. Traditional detection systems, which rely on static rules and predefined signatures, need more flexibility and adaptability to keep pace with these advancements. As a result, even minor changes in the structure or content of smishing messages can render these solutions ineffective, allowing a significant number of malicious messages to slip through undetected.

To bridge these critical gaps, this paper introduces a comprehensive approach aimed at substantially enhancing smishing attack detection capabilities. Recognizing the multifaceted challenges outlined above, we propose a novel content-based smishing detection framework that leverages advanced text normalization techniques to address the intricacies of SMS language. Our proposed framework employs a specialized Lingo Dictionary, meticulously curated to convert slang, abbreviations, and non-standard language into standardized forms. This normalization process is pivotal in reducing the linguistic variability that hampers traditional detection methods. By transforming diverse and often erratic language used in SMS into a consistent and uniform format, our approach ensures that the textual data is more easily interpretable by machine learning classifiers. Our approach significantly enhances the accuracy of machine learning classifiers in distinguishing between smishing and legitimate messages. This framework not only improves detection accuracy but also plays a crucial role in maintaining the integrity and security of users' personal information.\\

\noindent The main contributions of this paper are as follows:

\begin{itemize}
    \item \textit{\textbf{Enhanced Detection Model:}} We develop a smishing detection model that significantly improves the accuracy of identifying smishing messages compared to existing methods.
    
    \item \textit{\textbf{Advanced Text Normalization Techniques:}} We introduce sophisticated text normalization techniques into the smishing detection process, thereby enhancing the performance and effectiveness of classifiers by standardizing informal language, slang, and abbreviations commonly used in SMS communications.
    
    \item \textit{\textbf{Comprehensive Comparative Analysis:}} We conducted an extensive comparative analysis of our proposed smishing detection model, evaluating its performance against existing approaches using critical metrics such as precision, recall, and F1-score. By integrating normalization techniques with a Naïve Bayes classifier, our framework significantly improved accuracy from 88.2\% to 96.2\%. The True Positive Rate increased from 94.28\% to 97.14\%, and the True Negative Rate from 87.74\% to 96.12\%, outperforming existing methods. These results highlight the effectiveness of our approach in accurately distinguishing between legitimate and malicious messages.
\end{itemize}

The remainder of this paper is structured as follows: Section II discusses the background, including smishing statistics, attack procedures, techniques, and associated challenges. Section III reviews related work on SMS filtering techniques. Section IV details the proposed approach. Section V presents the experimental results, discussions, and a comparative analysis of our technique with existing smishing detection methods. Finally, Section VI concludes the paper and suggests directions for future research.

\section{Background}

The term “smishing", a blend of “SMS" and “phishing," emerged in the early 2000s as security researchers began uncovering vulnerabilities in text messaging services on various platforms, including Android and iOS \cite{infosec2018smishing}. As mobile communication grew in popularity, so did the exploitation of these vulnerabilities, establishing smishing as a significant cybersecurity threat. This section provides an overview of smishing attack procedures, techniques used by attackers, and the design goals and challenges in developing defences against these threats.

\subsection{Smishing Attack Procedure}

\begin{figure}[t!]
\centering
\includegraphics[width=0.45\textwidth]{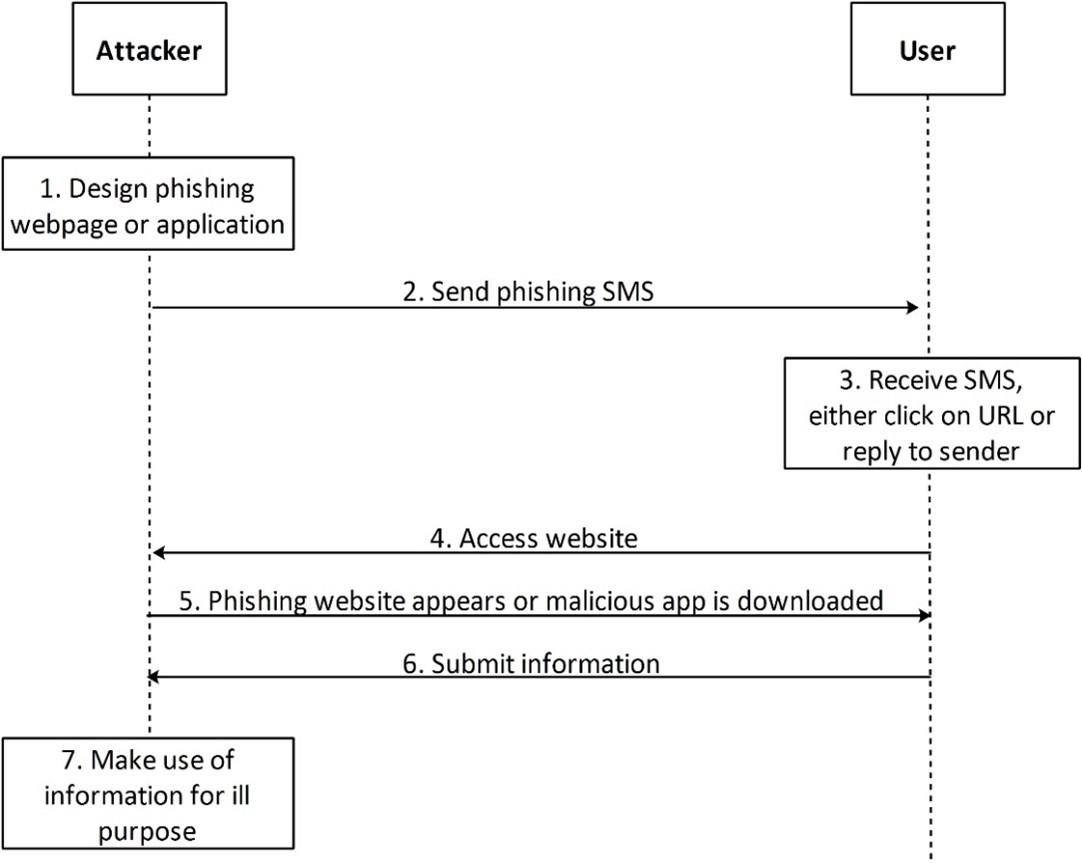}
\caption{Sequence Diagram of a Smishing Attack}
\label{fig1}
\end{figure}

A smishing attack is typically carried out through a series of carefully orchestrated steps aimed at deceiving users and extracting sensitive information, as illustrated in Figure \ref{fig1}.

\begin{enumerate}[leftmargin=*]
    \item \textit{{Design phishing webpage or application:}} The attacker creates a phishing webpage or a malicious application that closely mimics a legitimate site or app. This platform is designed with deceptive elements, including familiar logos, layouts, and branding, to trick users into believing they are interacting with a trusted source.
    
    \item \textit{{Send phishing SMS:}} The attacker sends an SMS to the targeted user containing a link to the phishing webpage or malicious application. The message often employs social engineering tactics, such as a sense of urgency or a warning, to increase the likelihood of user interaction.
    
    \item \textit{{Receive SMS, either click on URL or reply to sender:}} Upon receiving the SMS, the user reads the message and is prompted to take action by either clicking the provided URL or responding to the sender. This action initiates the attack process by directing the user to the phishing platform.
    
    \item \textit{{Access website:}} If the user clicks on the link, they are redirected to the phishing website. Attackers may use URL shortening services or domain obfuscation techniques to disguise the true destination of the link, making it harder for the user to recognize it as malicious.
    
    \item \textit{{Phishing website appears or malicious app is downloaded:}} Once redirected, the user encounters a phishing website that appears authentic or is prompted to download a malicious application. This website or app is designed to capture sensitive information or further compromise the device’s security.
    
    \item \textit{{Submit information:}} The user is then prompted to submit sensitive information, such as login credentials, personal identification details, or financial information. If the user complies, this information is directly transmitted to the attacker.
    
    \item \textit{{Make use of information for ill purpose:}} The attacker exploits the stolen information for malicious purposes, which may include unauthorized financial transactions, identity theft, or selling the information on the dark web. This misuse can result in significant harm to the victim, including financial loss and reputational damage.
\end{enumerate}

This process highlights the sophistication involved in smishing attacks, underscoring the need for robust detection and prevention mechanisms.

\subsection{Smishing Attack Techniques}

\begin{figure}[t!]
\centering
\includegraphics[width=0.45\textwidth]{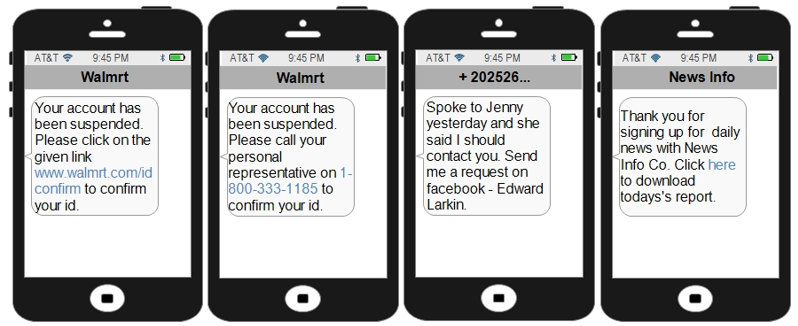}
\caption{Examples of Smishing Messages}
\label{fig2}
\end{figure}

Smishing attacks exploit specific vulnerabilities in user behaviour and trust through various techniques \cite{infosec2018smishing}, as illustrated in Figure \ref{fig2}. Key methods include:

\begin{enumerate}[leftmargin=*]
\item {\textit{Use of Bogus Links in SMS:}} Attackers send SMS messages that appear to be from reputable organizations, urging users to click a link to confirm their identity. This link leads to a phishing site designed to capture personal information.
\item {\textit{Use of Fake Phone Numbers:}} Users are asked to confirm their identity by calling a provided number, which connects them to a fraudulent call centre. Here, attackers use social engineering to extract sensitive information.
\item {\textit{Impersonation of Known Entities:}} Attackers impersonate individuals or entities familiar to the user, such as friends or colleagues, to establish trust and gather detailed information for further exploitation.
\item {\textit{Inducing the Download of Malicious Applications:}} Users are tricked into downloading a seemingly legitimate app via a link in the SMS. Once installed, the app secretly transmits the user's personal data to the attacker.
\end{enumerate}

These techniques demonstrate the adaptability and sophistication of smishing attackers, making it increasingly challenging for users and security systems to detect and mitigate these threats.

\begin{table*}[t!]
\centering
\caption{Summary of Related Work on Smishing Detection Techniques}
\label{tab:related_work}
\renewcommand{\arraystretch}{1.2}
\begin{tabular}{p{2.5cm}p{7cm}p{5.5cm}}
\hline
\textbf{Author} & \textbf{Approach} & \textbf{Key Results} \\ \hline

Yadav et al. \cite{yadav2011smsassassin} & 
“SMSAssassin" app using Naïve Bayes and SVM classifiers with crowd-sourced data. & 
97\% accuracy (ham), 72.5\% (spam) \\

Joo et al. \cite{joo2017s} & 
“S-Detector" with Naïve Bayes, analyzing URLs/APK files for smishing detection. & 
Real-time smishing blocking \\ 

Alfy et al. \cite{alfy2016spam} & 
Dendrite Cell Algorithm (DCA), inspired by the human immune system, filters SMS/emails. & 
Effective filtering on five datasets \\

Hauri Inc. \cite{smishing_defender} & 
"Smishing Defender" Android app for real-time detection/blocking with user alerts. & 
Real-time detection, further analysis option \\ 

Lee et al. \cite{lee2016smishing} & 
Cloud-based detection using source, content, and server location. & 
User-driven reduced false positives \\ 

Jain et al. \cite{jain2018rule} & 
Rule-based framework with nine content rules, tested on multiple algorithms. & 
High accuracy, best with RIPPER \\ 

Karami et al. \cite{karami2014improving} & 
Content-based approach with LIWC and SMS Specific (SMSS) features. & 
92\%-98\% accuracy \\ 

Silva et al. \cite{silva2017mdltext} & 
“MDLText," multinomial text classifier, minimum description length principle. & 
Low cost, robust against overfitting \\

Almeida et al. \cite{almeida2016text} & 
Text normalization using English/Lingo dictionaries, concept generation, and disambiguation. & 
Enhanced detection accuracy \\ 

Kaur et al. \cite{kaur2016normalization} & 
Hybrid text normalization combining machine translation/direct mapping. & 
98.4\% precision, 93.3\% recall, 90.1\% accuracy \\ \hline

\end{tabular}
\end{table*}

\subsection{Design Goals and Challenges}

Designing an effective smishing detection model for mobile devices involves addressing several vital challenges \cite{deepak2012challenging}:

\begin{enumerate}[leftmargin=*]
\item \textit{Optimized Computational Requirements:} Mobile devices have limited computational power and battery life, necessitating the development of lightweight algorithms that can perform real-time analysis without significantly impacting device performance.
\item \textit{Real-Time Detection and Response:} The model must quickly analyze incoming SMS messages, identify potential threats, and take immediate action, such as blocking or deleting the message, before the user interacts with it. This requires a delicate balance between speed and accuracy.
\item \textit{High Accuracy with Minimal False Positives/Negatives:} The model must accurately distinguish between legitimate messages and smishing attempts, minimizing both false positives (legitimate messages flagged as smishing) and false negatives (smishing messages classified as legitimate).
\item \textit{Preserving User Privacy:} The detection process should respect user privacy by analyzing messages locally on the device, avoiding the need to transmit sensitive content to external servers, which could pose a security risk.
\item \textit{Platform Independence and Compatibility:} The model should be platform-independent, functioning seamlessly across different mobile operating systems and devices despite variations in hardware and software environments.
\end{enumerate}

Meeting these goals is crucial for developing an effective and user-friendly smishing detection model that can be widely adopted to enhance mobile security.

\section{Related Work}
The detection and prevention of smishing attacks is a relatively nascent area of research. Yet, it has seen the development of various techniques aimed at addressing the unique challenges posed by smishing. Despite significant progress, there is still a considerable need for advanced methods. This section provides an overview of critical approaches in the literature, summarized in Table \ref{tab:related_work}.

Yadav et al. \cite{yadav2011smsassassin} developed SMSAssassin, an application for real-time spam filtering on mobile devices using Naïve Bayesian Classifier and Support Vector Machine (SVM). The system leverages crowd-sourcing to dynamically update spam patterns, achieving 97\% accuracy for ham messages and 72.5\% for spam on a dataset of 4,318 SMSes. Joo et al. \cite{joo2017s} introduced S-Detector, a smishing security framework that detects and blocks smishing messages while delivering legitimate ones. Their approach uses a Naïve Bayesian Classifier to analyze URLs in SMS messages, effectively filtering smishing content. Alfy et al. \cite{alfy2016spam} proposed a framework based on the Dendrite Cell Algorithm (DCA), which filters SMS and emails by mimicking the human immune system's response to threats. This machine learning-based approach was validated using five benchmark datasets. {Hauri} Inc. \cite{smishing_defender} developed Smishing Defender, an Android app providing real-time detection and blocking of smishing messages, with an option to submit suspicious messages for further analysis.

Lee et al. \cite{lee2016smishing} proposed a cloud-based smishing detection framework that processes message content in a virtual environment, allowing users to verify the legitimacy of messages. This user-involved process helps reduce false positives. Jain et al. \cite{jain2018rule} introduced a rule-based framework that identifies nine rules from message content to distinguish smishing from legitimate messages, demonstrating strong performance in zero-hour attack scenarios. Karami et al. \cite{karami2014improving} developed a content-based approach using semantic groupings of words, which enhances detection efficiency by reducing the feature set. Their system achieves accuracy rates between 92\% and 98\%. Silva et al. \cite{silva2017mdltext} created MDLText, a multinomial text classifier based on the minimum description length principle. This classifier is designed to be efficient, scalable, and resistant to overfitting, making it ideal for large-scale, dynamic environments. Almeida et al. \cite{almeida2016text} focused on text normalization, using English and Lingo dictionaries to standardize terms and resolve word ambiguity, which significantly improves detection accuracy. Kaur et al. \cite{kaur2016normalization} introduced a hybrid text normalization approach that combines statistical machine translation with direct mapping, achieving high precision, recall, and overall accuracy in translating short messages into standard English text.

\begin{figure*}[t!]
\centering
\includegraphics[width=0.65\paperwidth]{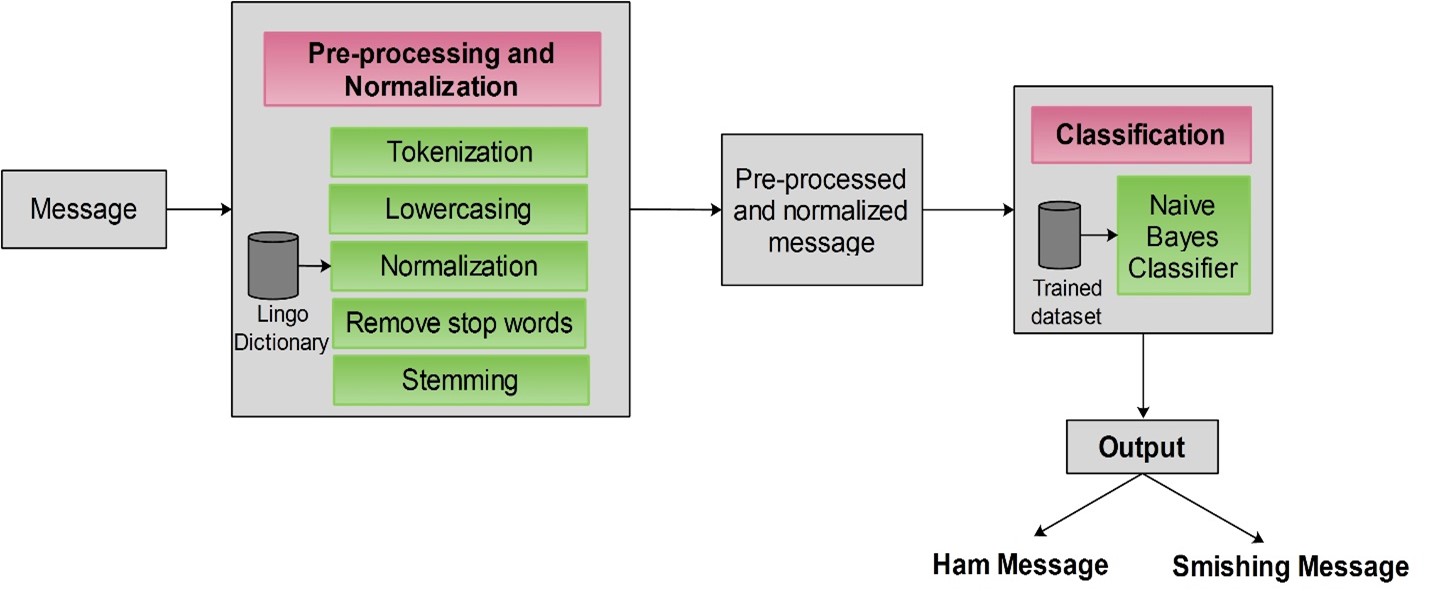}
\caption{Architecture of the Smishing Detection Framework.}
\label{fig3}
\end{figure*}

\section{Proposed Smishing Detection Framework}

In this section, we introduce our proposed smishing detection framework, designed to enhance the precision and effectiveness of identifying and mitigating smishing attacks. The architecture employs an advanced content-based analysis approach specifically tailored to address the linguistic challenges inherent in text messages, such as informal language, abbreviations, and slang. These linguistic variations can significantly undermine the accuracy of traditional classification methods. To overcome these challenges, our framework integrates a Lingo dictionary for text normalization, systematically converting non-standard language into its formal equivalents, thereby optimizing the accuracy of the subsequent classification process. The core classification mechanism utilizes the Naïve Bayesian Classifier, a well-established probabilistic model recognized for its robustness and efficiency in handling large-scale text classification tasks. The comprehensive architecture of the proposed framework is illustrated in Figure \ref{fig3}.

\begin{figure}[b!]
\centering
\includegraphics[width=0.95\columnwidth]{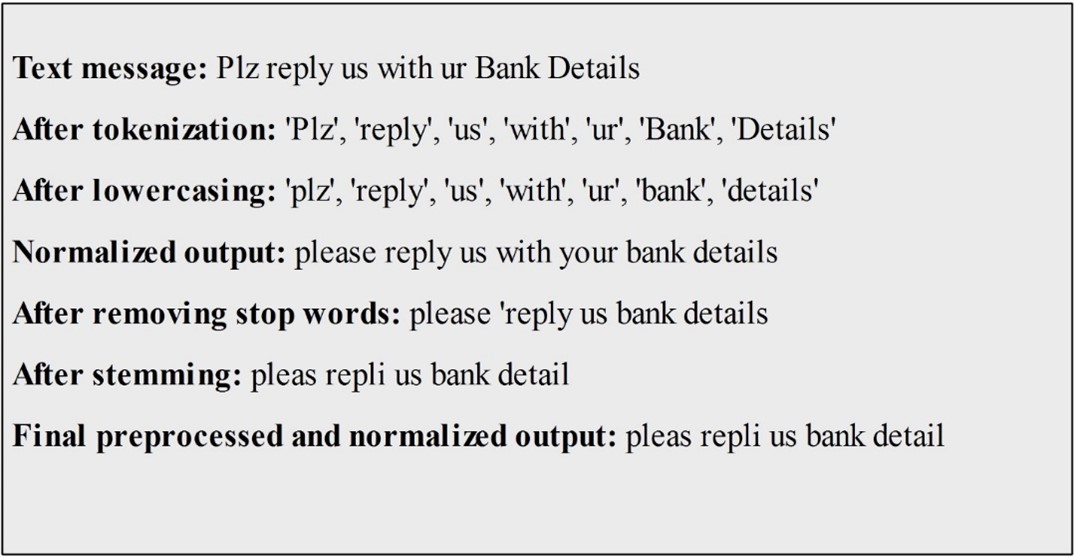}
\caption{Example of Preprocessing and Normalization.}
\label{fig4}
\end{figure}

\subsection{Phase 1: Preprocessing and Normalization}

The preprocessing and normalization phase serves as the cornerstone of our framework, meticulously designed to prepare incoming SMS messages for precise and reliable classification. This phase encompasses a sequence of carefully curated steps that systematically cleanse and standardize the text data, ensuring its suitability for subsequent analysis. The preprocessing phase includes several critical operations: tokenization, lowercasing, stop word removal, and stemming. These operations are succeeded by a rigorous normalization process, wherein slang, abbreviations, and other non-standard language forms are converted into their standard equivalents using the NoSlang dictionary \cite{noslang2017dictionary}. The outcome of this phase is a refined, normalized text dataset optimized for accurate classification in the subsequent stage. The core components of the preprocessing and normalization phase are as follows:

\begin{itemize}[leftmargin=*] 
\item \textit{\textbf{Tokenization:}} This process involves segmenting the SMS text into discrete tokens (words or terms), typically separated by whitespace or punctuation, thereby facilitating detailed syntactic and semantic analysis. 
\item \textit{\textbf{Lowercasing:}} All tokens are systematically converted to lowercase to ensure uniformity across the dataset, thereby mitigating potential complexities arising from case sensitivity in text analysis. 
\item \textit{\textbf{Normalization:}} This step entails replacing informal language, including slang and abbreviations, with their formal equivalents by leveraging the NoSlang dictionary. This standardization is essential for maintaining consistency in the subsequent classification process. 
\item \textit{\textbf{Stop Words Removal:}} Commonly occurring words that do not contribute to the meaningful classification of the text (e.g., articles, prepositions) are removed using the NLTK stop words list, thereby streamlining the text for more relevant feature extraction. 
\item \textit{\textbf{Stemming:}} Words are reduced to their root or base forms, eliminating linguistic variations and enhancing the classifier's ability to identify patterns within the text accurately. 
\end{itemize}

An example illustrating the preprocessing and normalization process is depicted in Figure \ref{fig4}. The detailed steps of this process are outlined in the Preprocessing and Normalization Algorithm, presented in Algorithm \ref{alog1}.

\begin{algorithm}[t!]
\caption{Preprocessing and Normalization Algorithm}
\label{alog1}
\begin{algorithmic}[1]
\Statex \textbf{Input:} msg (message), dict (NoSlang dictionary), stop (stop words)
\Statex \textbf{Output:} n\_msg (preprocessed and normalized message)
\State msg $\leftarrow$ msg.tokenization
\State msg $\leftarrow$ msg.lowercase
\For{each w in msg}
    \If{w found in dict}
        \State g $\leftarrow$ g.append (standard form)
    \Else
        \State g $\leftarrow$ g.append (w)
    \EndIf
\EndFor
\State msg $\leftarrow$ g
\For{each w in msg}
    \If{w found in stop}
        \State msg $\leftarrow$ msg.remove(w)
    \EndIf
\EndFor
\State msg $\leftarrow$ msg.stem
\State n\_msg $\leftarrow$ msg
\end{algorithmic}
\end{algorithm}

\subsection{Phase 2: Classification}

The classification phase constitutes the critical component of our proposed smishing detection framework, wherein preprocessed and normalized SMS messages are systematically categorized as either legitimate (ham) or malicious (smishing). This classification is performed using a Naïve Bayesian Classifier, a probabilistic model rooted in Bayes' theorem. The Naïve Bayesian Classifier is particularly well-suited for text classification tasks due to its capability to handle high-dimensional data and its robustness in managing noisy inputs.

Upon the completion of the preprocessing and normalization stages, each SMS message is transformed into a structured format that facilitates the classifier's analysis of linguistic patterns within the text. The Naïve Bayesian Classifier operates by calculating the posterior probability of each word in the message, thereby determining the likelihood that the message belongs to either the ham or smishing category. This is achieved through the use of a pre-trained dataset that has undergone identical preprocessing steps, ensuring consistency between the training and testing data. This methodology mitigates potential biases and significantly enhances classification accuracy.

Mathematically, the classification process is underpinned by Bayes' theorem, articulated as follows:

\begin{equation} p(C_k|x) = \frac{p(x|C_k)p(C_k)}{p(x)} \end{equation}

In this equation,  \( x \) denotes an attribute of the given data, \( p(C_k|x) \) is the posterior probability that data \( x \) belongs to class \( C_k \), \( p(x|C_k) \) represents the likelihood, and \( p(C_k) \) is the prior probability of class \( C_k \).

The classifier's performance is iteratively refined through rigorous training, as detailed in Algorithm \ref{alg:classification}. By integrating robust preprocessing and normalization techniques with precise classification using the Naïve Bayesian Classifier, the proposed framework significantly enhances the detection and blocking of smishing messages, thereby providing a formidable defence against this escalating cyber threat.

\begin{algorithm}[t!]
\caption{Classification Algorithm}
\label{alg:classification}
\begin{algorithmic}[1]
\Statex \textbf{Input:} D (dataset), n\_msg (preprocessed and normalized message)
\Statex \textbf{Output:} ham or smishing message
\State n\_Dataset $\leftarrow$ apply algorithm 1 on D
\State D\_train $\leftarrow$ select and extract 90\% of n\_Dataset
\State D\_test $\leftarrow$ select remaining messages of n\_Dataset
\For{each message m in D\_train}
    \For{each word w in message m}
        \State w\_ham $\leftarrow$ count number of ham messages in which w appears
        \State w\_smish $\leftarrow$ count number of smishing messages in which w appears
        \State w\_ham\_prob $\leftarrow$ (w\_ham) / total number of ham messages
        \State w\_smish\_prob $\leftarrow$ (w\_smish) / total number of smishing messages
    \EndFor
    \State ham\_train $\leftarrow$ ham probability of each word
    \State smish\_train $\leftarrow$ smish probability of each word
\EndFor
\State ham\_prob\_msg $\leftarrow$ apply equation 1 on n\_msg
\State smish\_prob\_msg $\leftarrow$ apply equation 1 on n\_msg
\If{(smish\_prob\_msg) > (ham\_prob\_msg)}
    \State output $\leftarrow$ smish message
\Else
    \State output $\leftarrow$ ham message
\EndIf
\State \textbf{return} output
\end{algorithmic}
\end{algorithm}

\section{Experimental Analysis and Discussion}

This section outlines the dataset used for the experimental analysis, the evaluation metrics employed, and the results that validate the effectiveness of our proposed framework.

\subsection{Dataset}

Effective experimental analysis relies on the quality and relevance of the dataset. For evaluating our proposed smishing detection framework, we utilized the SMS Spam Collection v.1 dataset \cite{almeida2016text}, a comprehensive collection of 5,574 English-language SMS messages. This dataset includes 4,827 ham (legitimate) messages and 747 spam messages. Given the absence of a dedicated smishing dataset, we pre-processed this dataset, manually extracting smishing messages from the spam category. Additionally, we enhanced our dataset by incorporating 71 smishing messages from Pinterest \cite{smishing2017pinterest}. The final dataset, consisting of 5,169 messages, i.e., 4,807 ham and 362 smishing, serves as the foundation for our experimental analysis. Descriptive statistics of this dataset is summarized in Table \ref{table:descriptive_stats}. On average, smishing messages are significantly longer, with 148.72 characters compared to 74.55 in ham messages, and contain more words, i.e., 24.72 versus 14.76. This verbosity likely reflects the attackers’ need to craft convincing fraudulent content. Additionally, URLs and financial symbols such as \$ and € are far more prevalent in smishing messages, with URLs appearing in 25.13\% of smishing messages compared to just 0.27\% of ham messages. Similarly, the occurrence of symbols like \$ and € is higher in smishing messages (1.93\%) than in ham messages (0.37\%). These descriptive insights highlight the distinct characteristics of smishing messages, which are critical for developing effective detection mechanisms. The findings from this dataset provide a solid foundation for the experimental evaluation and validation of our proposed smishing detection framework. Smishing detection is a binary classification task. The performance of the proposed framework is assessed using key metrics, as shown in Table \ref{table:performance_metrics}. Here, \( n_{\text{smish} \rightarrow \text{smish}} \) denotes the number of smishing messages correctly classified as smishing. Similarly,  \(  n_{\text{ham} \rightarrow \text{smish}} \) denotes the number of ham messages incorrectly classified as smishing.

\begin{table}[t!]
\centering
\caption{Descriptive Statistics of the Dataset}
\label{table:descriptive_stats}
\renewcommand{\arraystretch}{1.3}
\begin{tabular}{p{4.9cm}p{1.2cm}p{1.5cm}}
\hline
\textbf{} & {Ham Messages} & {Smishing Messages} \\ \hline
{Total Messages} & 4,807 & 362 \\ 
{Average No. of Characters} & 74.55 & 148.72 \\ 
{Average Presence of URLs} & 0.0027 & 0.2513 \\ 
{Average No. of Words} & 14.76 & 24.72 \\ 
{Average Presence of Symbols (\$ and €)} & 0.0037 & 0.0193 \\ \hline
\end{tabular}
\end{table}

\begin{table}[t!]
\centering
\caption{Performance Metrics Definitions}
\label{table:performance_metrics}
\renewcommand{\arraystretch}{1.5} 
\begin{tabular}{p{4cm}p{3cm}}
\hline
\textbf{Metric} & \textbf{Definition} \\ \hline
True Positives (TP) & $TP = n_{smish \rightarrow smish}$ \\ 
False Positives (FP) & $FP = n_{ham \rightarrow smish}$ \\ 
True Negatives (TN) & $TN = n_{ham \rightarrow ham}$ \\ 
False Negatives (FN) & $FN = n_{smish \rightarrow ham}$ \\ 
True Positive Rate (TPR) & $TPR = \frac{TP}{TP + FN}$ \\ 
False Positive Rate (FPR) & $FPR = \frac{FP}{TN + FP}$ \\ 
True Negative Rate (TNR) & $TNR = \frac{TN}{TN + FP}$ \\ 
False Negative Rate (FNR) & $FNR = \frac{FN}{FN + TP}$ \\ 
Accuracy (A) & $A = \frac{TP + TN}{TP + FP + FN + TN}$ \\ \hline
\end{tabular}
\end{table}

\subsection{Results and Discussions}

The proposed smishing detection framework was implemented on a computational setup featuring an Intel Core i5 processor (2.4 GHz) with 8 GB of RAM. The backend was developed in Python, utilizing various libraries essential for the framework's operation:

\begin{itemize}[leftmargin=*]
\item \textit{\textbf{Natural Language Processing Toolkit (NLTK):}} NLTK toolkit is used for text preprocessing and feature extraction, essential for the implementation of Natural Language Processing (NLP) tasks.
\item \textit{\textbf{CSV:}} CSV is employed to manage datasets in CSV format, enabling efficient data parsing, reading, and manipulation.
\item \textit{\textbf{SYS:}} It provides system-level functionalities, facilitating access to runtime environment variables and command-line argument management for streamlined script execution.
\item \textit{\textbf{ConfigParser:}} It managed configuration files dynamically, allowing the flexible configuration of parameters and paths within the codebase. The integration of OS modules ensured smooth interaction with the underlying operating system.
\end{itemize}

For the experimental evaluation, the dataset was divided into training and testing subsets, with 90\% of the data (4,342 ham messages and 327 smishing messages) allocated for training and the remaining 10\% reserved for testing. Both subsets underwent rigorous preprocessing and normalization, as detailed in Algorithm \ref{alog1}.


\subsubsection{Impact of Normalization on Detection Accuracy}

To assess the efficacy of the normalization process, we conducted a comparative analysis of smishing detection accuracy with and without normalization. The results are delineated in Tables \ref{table:trained_smishing} through \ref{table:trained_ham_normalization} and summarized in {Table} \ref{table:results_scheme}.\\

\noindent \textbf{Term Probability Analysis:}
\vspace{0.1in}

\begin{itemize}[leftmargin=*] \item \textbf{Without Normalization:} \begin{itemize} \item Smishing Dataset: Table \ref{table:trained_smishing} shows that the key terms such as Call (0.443425) and Claim (0.250764) exhibited high probabilities, indicating strong associations with smishing messages. However, the term Free also showed a substantial probability (0.159021) in smishing.

\begin{table}[t!]
\centering
\caption{Trained Smishing Dataset without Normalization}
\label{table:trained_smishing}
\renewcommand{\arraystretch}{1.1} 
\begin{tabular}{p{2cm}p{3cm}}
\hline
\textbf{Term} & \textbf{Smishing Probability} \\ \hline
Call & 0.443425 \\ 
Bank & 0.012232 \\ 
Cash & 0.159021 \\ 
Sale & 0.006116 \\ 
Offer & 0.033639 \\ 
Prize & 0.229358 \\ 
Free & 0.159021 \\ 
Won & 0.177370 \\ 
Claim & 0.250764 \\ \hline
\end{tabular}
\end{table}

\begin{table}[t!]
\centering
\caption{Trained Ham Dataset without Normalization}
\label{table:trained_ham}
\renewcommand{\arraystretch}{1.1} 
\begin{tabular}{p{2cm}p{3cm}}
\hline
\textbf{Term} & \textbf{Ham Probability} \\ \hline
Call & 0.062414 \\ 
Cash & 0.002303 \\ 
Sale & 0.001382 \\ 
Offer & 0.003685 \\ 
Offers & 0.000461 \\ 
Free & 0.030401 \\ 
Won & 0.005067 \\ 
Claim & 0.002533 \\ \hline
\end{tabular}
\end{table}

\begin{table}[t!]
\centering
\caption{Trained Smishing Dataset after Normalization}
\label{table:trained_smishing_normalization}
\renewcommand{\arraystretch}{1.1} 
\begin{tabular}{p{2cm}p{3cm}}
\hline
\textbf{Term} & \textbf{Smishing Probability} \\ \hline
Call & 0.464832 \\ 
Bank & 0.015291 \\ 
Cash & 0.159021 \\ 
Sale & 0.006116 \\ 
Offer & 0.055046 \\ 
Prize & 0.232416 \\ 
Free & 0.159021 \\ 
Won & 0.17737 \\ 
Claim & 0.253823 \\ \hline
\end{tabular}
\end{table}

\begin{table}[t!]
\centering
\caption{Trained Ham Dataset after Normalization}
\label{table:trained_ham_normalization}
\renewcommand{\arraystretch}{1.1} 
\begin{tabular}{p{2cm}p{3cm}}
\hline
\textbf{Term} & \textbf{Ham Probability} \\ \hline
Call & 0.071165 \\ 
Bank & 0.002303 \\ 
Cash & 0.002533 \\ 
Sale & 0.001842 \\ 
Offer & 0.004376 \\ 
Free & 0.030631 \\ 
Won & 0.005067 \\ 
Claim & 0.002994 \\ \hline
\end{tabular}
\end{table}

\item Ham Dataset: Table \ref{table:trained_ham} shows that terms like Call (0.062414) and Free (0.030401) had lower probabilities, reflecting their limited use in legitimate messages. Hence, they are mostly used in smishing messages. \end{itemize}

\item \textbf{After Normalization:}
\begin{itemize}
    \item Smishing Dataset: Table \ref{table:trained_smishing_normalization} shows that Post-normalization the probability of ``Call'' increased to 0.464832, and ``Claim'' rose to 0.253823, reinforcing their association with smishing. The normalization process enhanced the consistency of term usage, thereby improving the model's discriminative capability.
    \item Ham Dataset: The probability of ``Call'' in ham messages increased slightly to 0.071165, while other terms showed minimal adjustments. This subtle shift aids in better distinguishing between legitimate and malicious messages.
\end{itemize}

\end{itemize}

\begin{table}[t!]
\centering
\caption{Results Comparison of Proposed Approach Before and After Normalization}
\label{table:results_scheme}
\renewcommand{\arraystretch}{1.1} 
\begin{tabular}{p{3cm}p{2cm}p{2cm}}
\hline
\textbf{Metric} & \textbf{Without Normalization} & \textbf{With Normalization} \\ \hline
Accuracy & 88.2\% & 96.2\% \\ 
True Positive Rate (TPR) & 94.28\% & 97.14\% \\ 
True Negative Rate (TNR) & 87.74\% & 96.12\% \\ \hline
\end{tabular}
\end{table}

\begin{table}[t!]
\centering
\caption{Comparative Analysis with Existing Approaches}
\label{table:comparison_existing}
\renewcommand{\arraystretch}{1.1} 
\begin{tabular}{p{3cm}p{1cm}p{1cm}p{1.5cm}}
\hline
\textbf{Approach} & \textbf{Accuracy} & \textbf{TPR} & \textbf{TNR} \\ \hline
Baseline 1 & 85.6\% & 91.3\% & 84.2\% \\ 
Baseline 2 & 87.4\% & 93.0\% &85.6\% \\ 
\textbf{Proposed Framework (with Normalization)} & \textbf{96.2\%} & \textbf{97.14\%} & \textbf{96.12\%} \\ \hline
\end{tabular}
\end{table}

\vspace{0.1in}
\noindent \textbf{Performance Metrics:} The classification performance was evaluated using key metrics: Accuracy, True Positive Rate (TPR), and True Negative Rate (TNR). The comparative results are presented in Table \ref{table:results_scheme}. The results demonstrate a significant improvement across all metrics post-normalization:

\begin{itemize}[leftmargin=*] 
\item \textit{Accuracy}: Increased from 88.2\% to 96.2\%, indicating a substantial enhancement in the framework's overall ability to classify messages correctly. 
\item \textit{True Positive Rate (TPR)}: Improved from 94.28\% to 97.14\%, reflecting a higher sensitivity in detecting smishing messages. 
\item \textit{True Negative Rate (TNR):} Elevated from 87.74\% to 96.12\%, showcasing a marked improvement in correctly identifying legitimate (ham) messages. \end{itemize}

\vspace{0.1in}

\noindent \textbf{Comparative Analysis with Existing Approaches:} Table \ref{table:comparison_existing} presents a comparative analysis between the proposed smishing detection framework and existing approaches. The results clearly demonstrate that our framework significantly outperforms the existing methods across all evaluated metrics. Specifically, the proposed framework achieves an {accuracy} of {96.2\%}, a TPR of {97.14\%}, and a TNR of {96.12\%}. In contrast, Existing Approach 1 and Existing Approach 2 attain accuracies of 85.6\% and {87.4\%}, TPRs of {91.3\%} and {93.0\%}, and TNRs of {84.2\%} and {85.6\%}, respectively. The substantial improvements in accuracy, TPR, and TNR underscore the effectiveness of incorporating normalization within our framework, as well as the robustness of the Naïve Bayesian Classifier in accurately distinguishing between legitimate and malicious messages. These enhancements collectively contribute to a more reliable and efficient smishing detection system, setting a new benchmark in the field.

Table \ref{table:results_scheme} showcases the performance of the proposed smishing detection framework with and without normalization. Incorporating normalization leads to substantial improvements across all metrics: the TPR increases from {94.28\%} to {97.14\%}, and the TNR rises sharply from 87.74\% to 96.12\%. Concurrently, the FPR and FNR decrease significantly from {12.25\%} to {3.87\%} and {5.71\%} to {2.85\%}, respectively. Overall, the {Accuracy} of the framework enhances from {88.2\%} to {96.2\%}, underscoring the critical role of normalization in boosting the model's precision and reliability in detecting smishing threats.

Table \ref{table:comparison_existing} illustrates the comparison between our proposed smishing detection framework and existing approaches. The results clearly demonstrate that our framework, which integrates content-based analysis and normalization using the Naive Bayes algorithm, achieves superior performance metrics.

\begin{table*}[!t]
\centering
\caption{Experimental Results of the Scheme}
\label{table:results_scheme}
\renewcommand{\arraystretch}{1.4} 
\begin{tabular}{cccccc}
\hline
\textbf{} & \textbf{TPR} & \textbf{TNR} & \textbf{FPR} & \textbf{FNR} & \textbf{Accuracy} \\ \hline
Without normalization & 94.28\% & 87.74\% & 12.25\% & 5.71\% & 88.2\% \\ \hline
With normalization & 97.14\% & 96.12\% & 3.87\% & 2.85\% & 96.2\% \\ \hline
\end{tabular}
\end{table*}

\begin{table*}[!htbp]
\centering
\caption{Comparison of Our Proposed System with Existing Approaches}
\label{table:comparison_existing}
\renewcommand{\arraystretch}{1.2} 
\begin{tabular}{p{3cm}ccp{3cm}p{5cm}}
\hline
\textbf{Authors} & \textbf{Content-based Analysis} & \textbf{Normalization} & \textbf{Algorithm Used} & \textbf{Major Findings} \\ \hline
Yadav et al. \cite{yadav2011smsassassin} & \checkmark & \xmark & Naive Bayes Algorithm, SVM & Achieved 97\% classification accuracy for ham messages and 72.5\% for spam messages. \\ \hline
Joo et al. \cite{joo2017s} & \checkmark & \xmark & Naive Bayes Algorithm & Utilized statistical learning methods to identify words frequently used in smishing text messages. \\ \hline
Smishing Defender \cite{smishing_defender} & \checkmark & \xmark & - & Developed a real-time application that can be downloaded and installed for immediate evaluation. \\ \hline
Lee et al. \cite{lee2016smishing} & \checkmark & \xmark & Source analysis, content analysis, server location determination & Processing conducted in a cloud environment. User involvement reduces the false detection rate. \\ \hline
Proposed Framework without Normalization & \checkmark & \xmark & Naive Bayes Algorithm & Achieved 88.2\% accuracy with 94.28\% TPR and 87.74\% TNR. \\ \hline
Proposed Framework with Normalization & \checkmark & \checkmark & Naive Bayes Algorithm & Achieved 96.2\% accuracy with 97.14\% TPR and 96.12\% TNR. \\ \hline
\end{tabular}
\end{table*}

\section{Conclusion and Future Work}

The widespread adoption of smartphone devices has concurrently led to an increase in related security threats, with smishing attacks emerging as a prominent concern for mobile phone users. Accurately identifying smishing attacks remains a critical research challenge, particularly given the limited existing work in this domain. Although various detection solutions have been proposed, a comprehensive and effective solution still needs to be improved. A significant challenge is the frequent use of abbreviations, slang, and short forms in messages, which can cause ambiguous word interpretations. Additionally, the concise nature of these messages restricts the number of extractable features for analysis. To address these challenges, we developed a smishing detection model that preprocesses messages and normalizes them using the NoSlang dictionary. We used Bayesian learning techniques to train the dataset, resulting in separate training sets for legitimate (ham) and smishing messages. A Naïve Bayesian classifier is developed to classify messages as either smishing or ham. We evaluate our approach using a publicly available dataset, and the experimental results demonstrate that the normalization process significantly enhances the classifier's performance. The proposed method achieves high accuracy, TPR, and TNR of 96.2\%, 97.14\%, and 96.12\%, respectively.

Future work will focus on improving the normalization process to expand message words based on their contextual relevance from related concepts. Additionally, extending the dataset with more messages will enhance the model’s robustness and effectiveness. We also plan to analyze URLs embedded within messages to determine whether they lead to the download of malicious applications or redirect users to fraudulent login pages. These enhancements aim to strengthen the detection capabilities further and provide a more comprehensive solution to combating smishing attacks.

\end{document}